\documentclass[pra,showpacs,tightenlines,twocolumn,floatfix]{revtex4}

\usepackage{graphicx}  %
\usepackage{bm}  %
\usepackage{amssymb}
\usepackage{dcolumn}

\begin{document}

%


\newcommand{\beq}{\begin{equation}}
\newcommand{\eeq}{\end{equation}}
\newcommand{\bea}{\begin{eqnarray}}
\newcommand{\eea}{\end{eqnarray}}
\newcommand{\ben}{\begin{eqnarray*}}
\newcommand{\een}{\end{eqnarray*}}

\newcommand{\xvec}{\vector{\bf{x}}}
\newcommand{\zvec}{\vector{z}}
\newcommand{\rvec}{\vector{r}}

\newcommand{\simlt}{\stackrel{<}{{}_\sim}}
\newcommand{\simgt}{\stackrel{>}{{}_\sim}}
\newcommand{\sing}{$^1\!S_0$ }
\newcommand{\btau}{\mbox{\boldmath$\tau$}}
\newcommand{\bsig}{\mbox{\boldmath$\sigma$}}

\newcommand{\dt}{\partial_t}

\newcommand{\kf}{k_{\rm F}}
\newcommand{\wt}{\widetilde}
\newcommand{\kt}{\widetilde k}
\newcommand{\pt}{\widetilde p}
\newcommand{\qt}{\widetilde q}
\newcommand{\wh}{\widehat}
\newcommand{\dens}{\rho}
\newcommand{\edens}{{\cal E}}
\newcommand{\order}[1]{{\cal O}(#1)}

\newcommand{\psihat}{\widehat\psi}
\newcommand{\dagphan}{{\phantom{\dagger}}}
\newcommand{\kvec}{{\bf k}}
\newcommand{\kpvec}{{\bf k}'}
\newcommand{\ak}{a^\dagphan_\kvec}
\newcommand{\akdag}{a^\dagger_\kvec}
\newcommand{\akv}[1]{a^\dagphan_{\kvec_{#1}}}
\newcommand{\akdagv}[1]{a^\dagger_{\kvec_{#1}}}
\newcommand{\akp}{a^\dagphan_{\kvec'}}
\newcommand{\akpdag}{a^\dagger_{\kvec'}}
\newcommand{\akpv}[1]{a^\dagphan_{\kvec'_{#1}}}
\newcommand{\akpdagv}[1]{a^\dagger_{\kvec'_{#1}}}

\def\vec#1{{\bf #1}}

\newcommand{\nab}{\overrightarrow{\nabla}}
\newcommand{\nabsq}{\overrightarrow{\nabla}^{2}\!}
\newcommand{\nabl}{\overleftarrow{\nabla}}
\newcommand{\galnab}{\tensor{\nabla}}
\newcommand{\psid}{{\psi^\dagger}}
\newcommand{\psidal}{{\psi^\dagger_\alpha}}
\newcommand{\psidbe}{{\psi^\dagger_\beta}}
\newcommand{\idt}{{i\partial_t}}
\newcommand{\Sthree}{{\delta_{11'}(\delta_{22'}\delta_{33'}%
        -\delta_{23'}\delta_{32'})%
        +\delta_{12'}(\delta_{23'}\delta_{31'}-\delta_{21'}\delta_{33'})%
        +\delta_{13'}(\delta_{21'}\delta_{32'}-\delta_{22'}\delta_{31'})}}
\newcommand{\Stwo}{{\delta_{11'}\delta_{22'}-\delta_{12'}\delta_{21'}}}
\newcommand{\Left}{{\cal L}}
\newcommand{\Tr}{{\rm Tr}}

\newcommand{\h}{\hfil}
\newcommand{\be}{\begin{enumerate}}
\newcommand{\ee}{\end{enumerate}}
\newcommand{\I}{\item}   

\newcommand{\density}{\rho}

\newcommand{\thyp}{\mbox{---}}

\newcommand{\Jks}{J_{\ks}}
\newcommand{\Jzero}{\Jks}
\newcommand{\Jdensityzero}{J_\density^0}

\newcommand{\ks}{{\rm ks}}
\newcommand{\Seq}{Schr\"odinger\ equation }
\newcommand{\Rvec}{{\bf R}}
\newcommand{\yvec}{{\bf y}}
\newcommand{\ve}{V_{eff}}
\newcommand{\densityJzero}{\density^0_J}
\newcommand{\Dfunct}{{D^{-1}}}
\newcommand{\drv}[2]{{\mbox{$\partial$} #1\over \mbox{$\partial$} #2}}
\newcommand{\drvs}[2]{{\mbox{$\partial^2$} #1\over \mbox{$\partial$} #2 \mbox{$^2$}}}
\newcommand{\drvt}[2]{{\partial^3 #1\over \partial #2 ^3}}
\newcommand{\til}[1]{{\widetilde #1}}
\newcommand{\dthreex}{d^3\xvec}
\newcommand{\dthreey}{d^3\yvec}

\newcommand{\efermi}{\varepsilon_{{\scriptscriptstyle \rm F}}}
\newcommand{\eHF}{\wt\varepsilon}
\newcommand{\eps}{\varepsilon}
\newcommand{\eKS}{e}
\newcommand{\ekJ}{e_\kvec^J}
\newcommand{\epsk}{\varepsilon_\kvec}
\newcommand{\epsKS}{\varepsilon}
\newcommand{\Eq}[1]{Eq.~(\ref{#1})}

\newcommand{\Fi}[1]{\mbox{$F_{#1}$}}
\newcommand{\fq}{f_{\qvec}}

\newcommand{\Gammaalt}{\overline\Gamma}
\newcommand{\Gammaks}{\Gamma_{\ks}}
\newcommand{\Gamint}{\widetilde{\Gamma}_{\rm int}}
\newcommand{\GKS}{G_{\ks}}
\newcommand{\grad}{{\bm{\nabla}}}   
\newcommand{\greenKS}{{G}_{\ks}}
\newcommand{\intint}{\int\!\!\int}

\newcommand{\kfermi}{k_{{\scriptscriptstyle \rm F}}}   
\newcommand{\kJzero}{k_J}

\renewcommand{\l}{\lambda}

\newcommand{\meV}{\mbox{\,meV}}
\newcommand{\mi}[1]{\mbox{$\mu_{#1}$}}

\newcommand{\Oi}[1]{\mbox{$\Omega_{#1}$}}

\newcommand{\phibar}{\overline\phi}
\newcommand{\phidagger}{\phi^\dagger}
\newcommand{\phistar}{\phi^\ast}
\newcommand{\psibar}{\overline\psi}
\newcommand{\psidagger}{\psi^\dagger}
\newcommand{\qvec}{\vector{\rho}}

\newcommand{\tr}{{\rm tr\,}}

\newcommand{\Ulong}{U_{L}}

\renewcommand{\vector}[1]{{\bf #1}}
\newcommand{\vext}{v_{\rm ext}}   
\newcommand{\Vlong}{V_{L}}

\newcommand{\Wzero}{W_0}
\newcommand{\Wks}{W_{\ks}}

\newcommand{\ts}{\textstyle}
%
%
%
%


%
\title{Density functional theory for fermions close to the unitary regime}
\author{Anirban Bhattacharyya}\email{anirban@ornl.gov}
\author{T. Papenbrock}\email{papenbro@phy.ornl.gov}
\affiliation{
Department of Physics and Astronomy, University of Tennessee, 
Knoxville, TN\ 37996, and\\
Physics Division, Oak Ridge National Laboratory, Oak Ridge, TN\ 37831
}

\date{\today}
%

\begin{abstract}
%
We consider interacting Fermi systems close to the unitary regime and compute
the corrections to the energy density that are due to a large scattering 
length and a small effective range. Our approach exploits the universality
of the density functional and determines the corrections from the analyical
results for the harmonically trapped two-body system. 
The corrections due to the finite scattering length compare well with the 
result of Monte Carlo simulations. We also apply our results to symmetric 
neutron matter.
\end{abstract}
\maketitle

Ultracold fermionic atom gases have attracted a lot of interest since
Fermi degeneracy was achieved by DeMarco and Jin \cite{DeM99}. These
systems are in the metastable gas phase, as three-body recombinations
are rare. Most interestingly, the effective two-body interaction
itself can be controlled via external magnetic fields.  This makes it
possible to study the system as it evolves from a dilute Fermi gas
with weak attractive interactions to a bosonic gas of diatomic
molecules. This transition from a superfluid BCS state to Bose
Einstein condensation (BEC) has been the subject of many experimental
\cite{OHa02,Reg03,Bou03,Zwi03,Geh03,Reg04,Kin04,Bar04,Bou04,Gre05,Kin05}
and theoretical works
\cite{Hol01,Men02,Bul03,Kim04,Bul05,Ho03,Bul05b,Sch05,Car03,Ast04,Hei01}.

At the midpoint of this transition, the two-body system has a
zero-energy bound state, and the scattering length diverges.  If
other parameters as the effective range of the interaction can be
neglected, the interparticle spacing becomes the only relevant length
scale. This defines the unitary limit. In this limit, the energy
density is proportional of that of a free Fermi gas, the
proportionality constant denoted by $\xi$. Close to the unitary limit, 
corrections are due to a finite, large scattering length $a$ and a
small effective range $r_0$ of the potential. Within the local density
approximation (LDA), the energy density is given as
\beq
\label{lda}
{\cal E}[\rho] = {\cal E}_{\rm FG}
\left(\xi  + {c_1\over a \rho^{1/3}} + c_2 r_0 \rho^{1/3}\right). 
\eeq
Here, 
\beq
\label{FG}
{\cal E}_{\rm FG}[\rho] = {3\over 10}\left(3\pi^2\right)^{2/3}
{\hbar^2\over m} \rho^{5/3} 
\eeq 
is the energy density of the free Fermi gas. The universal constant
$\xi$ has been computed by several authors. Monte Carlo calculations
by Carlson {\it et al.}  \cite{Car03}, Astrakharchik {\it et al.}
\cite{Ast04}, and by Bulgac {\it et al.} \cite{Bul05b} agree well with
each other and yield $\xi\approx 0.44\pm 0.01$, $\xi\approx 0.42\pm
0.01$, and $\xi\approx 0.42$, respectively. A calculation by Steele
\cite{Ste00} based on effective field theory yields $\xi=4/9$, while
an application of density functional theory (DFT) \cite{HK64,KOHN65}
yields $\xi\approx 0.42$ \cite{Pap05}. Other calculations deviate
considerably from these results. Heiselberg \cite{Hei01} obtained
$\xi=0.326$, while Baker \cite{Bak99} found $\xi=0.326$ and
$\xi=0.568$ from different Pad{\'e} approximations to Fermi gas
expansions.  Engelbrecht {\it et al.}  \cite{Eng97} obtained
$\xi=0.59$ in a calculation based on BCS theory, while a very recent
Monte Carlo simulation by Lee \cite{Lee05} yields $\xi\approx 0.25$.
The experimental values are $\xi\approx 0.74\pm 0.07$~\cite{Geh03},
$\xi= 0.51\pm 0.04$~\cite{Kin05}, $\xi\approx 0.7$~\cite{Bou03}, $\xi=
0.27^{+0.12}_{-0.09}$~\cite{Bar04}. The constant $c_1$ in
Eq.~(\ref{lda}) has also been determined. The Monte Carlo results by
Chang {\it et al.} \cite{Cha04} and by Astrakharchik {\it et al.}
\cite{Ast04} yield $c_1\approx - 0.28$~\cite{BB05} 
and are very close to Steele's
analytical result \cite{Ste00}.  We are not aware of any estimate for
the constant $c_2$ in Eq.~(\ref{lda}) that concerns the correction due
to a small effective range. It is the purpose of this work to fill
this gap. This is particularly interesting as experiments also have
control over the effective range. Note that the regime of a large
effective range has recently been discussed by Schwenk and Pethick
\cite{Sch05}.

In this work, we determine the coefficients $c_1$, and $c_2$ via
density functional theory. Recall that the density functional is
supposed to be universal, i.e. it can be used to solve the $N$-fermion
system for any particle number $N$, and for any external potential.
Exploiting the universality of the density functional, the parameters
$c_1$ and $c_2$ can be obtained from a fit to an analytically known
solution, i.e. the harmonically trapped two-fermion system
\cite{Bus98}. This simple approach has recently been applied
\cite{Pap05} to determine the universal constant $\xi$, and will be 
followed and extended below.

Let us briefly turn to the harmonically trapped two-fermion system.
The wave function $u(r)$ in the relative
coordinate $r=r_1-r_2$ of the spin-singlet state is given in terms
of the parabolic cylinder function $U(-\varepsilon, r/\lambda)$
\cite{Bus98,Block,Abramowitz}. Here, $\varepsilon\hbar\omega$ is the
relative energy, and $\lambda=\sqrt{\hbar/(m\omega)}$ denotes
the oscillator length. We are dealing with a short-ranged two-body
interaction and quantize the energy through the boundary condition at
the origin
\beq
\label{quant}
{\partial_r u(r)\over u(r)}\bigg|_{r=0} = k \cot{\delta},
\eeq
where $\hbar^2k^2/m=\varepsilon\hbar\omega$, and $\delta$ denotes
the $s$-wave phase shift. The evaluation of Eq.~(\ref{quant}) for the
parabolic cylinder function yields
\beq
\label{master}
\sqrt{2}\,{\Gamma(3/4-\varepsilon/2)\over\Gamma(1/4-\varepsilon/2)} 
= {\lambda\over a} - {r_0\varepsilon\over 2\lambda}.
\eeq
Here, we have employed the effective range expansion of the phase
shift. Note that Eq.~(\ref{master}) is valid for arbitrary values of
the scattering length $a$ and the effective range $r_0$.

As an introductory example, we consider the case of a dilute Fermi gas
with a small value of the (positive) scattering length $a\ll \lambda$
and zero range. We expand Eq.~(\ref{master}) around the energy of the
noninteracting system as $\varepsilon = 3/2 + \Delta\varepsilon$. The
energy correction fulfills $\Delta\varepsilon \ll 1$, and we find
\beq
\label{Esmalla}
\Delta\varepsilon = \sqrt{2\over \pi} {a\over \lambda}.
\eeq
The form of this result suggests that the energy density of the weakly
interacting system is that of the noninteracting system plus the
term
\beq
\label{smalla}
\Delta{\cal E}[\rho]=c \left(a\rho^{1/3}\right) {\hbar^2\over m}\rho^{5/3}, 
\eeq
which is due to the scattering length.  We want to determine the
coefficient $c$ in Eq.~(\ref{smalla}). Recall that Kohn-Sham DFT is
variational, and that we are dealing with a small perturbation $a\ll
\rho^{1/3}$.  Thus, we can insert the density of the noninteracting
system $\rho(r)=2\pi^{-3/2} \lambda^{-3} e^{-r^2/\lambda^2}$ into
Eq.~(\ref{smalla}) and integrate over all space. Equating the result
with the energy correction given by Eq.~(\ref{Esmalla}) yields
$c=\pi$, which is in full agreement with many-body perturbation theory
\cite{Lenz,Huang,HAmmER00}.  This result is not really surprising. The
interaction is a contact interaction, and the energy correction given
by Eq.~(\ref{smalla}) is the Hartree-Fock approximation of this
interaction. Nevertheless, it is encouraging that the simple DFT
approach via the two-body system yields a result in agreement with
many-body theory.

\begin{figure}
\centerline{\includegraphics*[width=6cm,angle=-90]{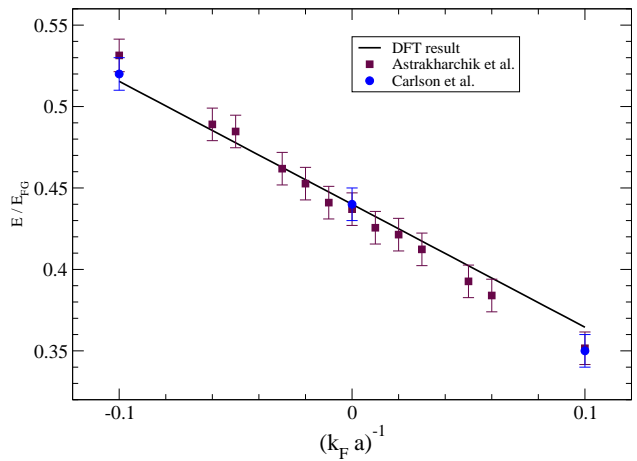}}
\vspace*{-.1in}
\caption{Energy per particle (in units of the free Fermi gas)  
as a function of $(\kf a)^{-1}$ in the vicinity of the unitary regime. 
Solid line: slope estimated in this work; data points: Monte Carlo results
from Ref.~\cite{Car03} (dots) and from Ref.~\cite{Ast04} (squares), 
respectively.}
\label{fig1}
\end{figure}

Let us turn to the vicinity of the unitary regime.  Consider the case
of a large scattering length $a\gg \lambda$ and zero range. We expand
Eq.~(\ref{master}) around the energy corresponding to the unitary
regime as $\varepsilon=1/2 +\Delta\varepsilon$, and find
\beq
\label{Elargea}
\Delta\varepsilon = -\sqrt{2\over\pi}{\lambda\over a}.
\eeq
This expression suggests that the correction to  
the energy density $\xi {\cal E}_{\rm FG}$ is of the form  
\beq
\label{largea}
\Delta{\cal E}_1[\rho] = {c_1\over a\rho^{1/3}}\,{\cal E}_{\rm FG}[\rho].
\eeq
We insert the exact density at the unitary regime, 
\beq
\label{dens}
\rho(r)={4e^{-2(r/\lambda)^2}\over \pi^{3/2} \lambda^2 r}
\int\limits_0^{r/\lambda} dx \, e^{x^2}
={2e^{-2r^2/\lambda^2}\over \pi\lambda^2 r} {\rm Erfi}(r/\lambda)
\eeq
into the correction given by Eq.~(\ref{largea}) and
integrate. Equating the result with the exact result~(\ref{Elargea}) 
yields $c_1= -0.244$.  Monte Carlo calculations
predict $c_1\approx -0.28$. Our result deviates only 13\% from the
results of the Monte Carlo calculations (see Fig.~\ref{fig1}). The
deviation is due to the fact that the simple functional in
Eq.~(\ref{largea}) is the LDA of the (unknown) exact density
functional. Given the simplicity of our approach, the estimate is
remarkably accurate.

Let us consider the corrections due to a non-zero effective range
$r_0\ll \lambda$. Again, we expand Eq.~(\ref{master}) around the energy of
the unitary regime as $\varepsilon=1/2 +\Delta\varepsilon$, and find
\beq
\label{Erange}
\Delta\varepsilon = {1\over\sqrt{8\pi}} {r_0\over\lambda}.
\eeq
The form of this energy correction implies that the term
\beq
\label{range}
\Delta{\cal E}_2[\rho] = c_2 r_0\rho^{1/3}\, {\cal E}_{\rm FG}[\rho]
\eeq
has to be added to the energy density $\xi {\cal E}_{\rm FG}$.  For a
determination of the coefficient $c_2$, we insert the density given by
Eq.~(\ref{dens}) in Eq.~(\ref{range}) and integrate. Comparison of the
result with the exact result~(\ref{Erange}) yields $c_2=0.142$. This
is one of the main results of this work. We estimate that the
systematic error of this coefficient is about 5\%-15\%, as this is the
deviation by which the DFT estimates for $\xi$ \cite{Pap05}, and $c_1$
deviates from the Monte Carlo predictions \cite{Car03,Ast04}. The
estimate for $c_2$ enables us to discuss a small systematic correction
of the universal constant obtained from Monte Carlo calculations.
Recall that the Monte Carlo calculations \cite{Car03} and \cite{Ast04}
are based on potentials with a small effective range of about
$r_0\rho^{1/3}\approx 0.05$, and $r_0\rho^{1/3} = 0.01$,
respectively. This suggests that their predictions for the universal
constant $\xi$ involve a small positive error of about
$c_2r_0\rho^{1/3}\approx 0.007$ and $c_2r_0\rho^{1/3}\approx 0.001$,
respectively, which is within the statistical error of these
simulations.

We also tried to improve the accuracy of our estimates for $c_1$ and
$c_2$ by going beyond the LDA. The main idea consists of adding 
gradient terms to the energy functional, and to use Kohn-Sham DFT.
The systematic inclusion of the nonlocal kinetic energy density
in the energy functional can lead to improvements in the density
and energy spectrum \cite{FURNSTAHL04b,AB3}. Here, we follow a
phenomenological approach. We replace the functional in 
Eq.~(\ref{lda}) by the functional  
\beq
\label{ks}
{\cal E}[\rho] = \xi {\cal E}_{\xi}[\rho] + {c_1\over a\rho^{1/3}}
{\cal E}_{a}[\rho] + c_2(r_0\rho^{1/3}){\cal E}_{r_0}[\rho]. 
\eeq
Here 
\beq
\label{grad}
{\cal E}_{\xi}[\rho] = {\hbar^2\over m}\left({f_\xi\over 2}\sum_{j=1}^N
|\grad\phi_j|^2 + 
(1-f_\xi){3\over 10}(3\pi^2)^{2\over 3}\rho^{5\over 3}\right), 
\eeq
and similar expressions with parameters $f_a$ and $f_{r_0}$ are
employed for the terms involving the scattering length and the
effective range, respectively. Note that the functional~(\ref{lda}) is
the Thomas-Fermi approximation of the functional~(\ref{ks}), and that
both functionals are identical for $f_\xi=f_a=f_{r_0}=0$. Note also
that the density-dependent term in Eq.~(\ref{grad}) is the
Thomas-Fermi approximation of the corresponding gradient term.  The
pair of parameters $(\xi,f_\xi)$ was determined in Ref.~\cite{Pap05},
and the universal constant $\xi$ varies only very little when $f_\xi$
is varied.  This is very different for the parameter pairs $(c_1,f_a)$
and $(c_2,f_{r_0})$, as the energy obtained from integration of the
gradient term in Eq.~(\ref{grad}) differs by a factor of 2.1 and 0.7,
respectively from the energy of the corresponding density-dependent
term.  This finding indicates that the functionals ${\cal E}_a[\rho]$
and ${\cal E}_{r_0}[\rho]$ exhibit considerable finite-size
corrections (as the gradient terms differ from their respective 
Thomas-Fermi limits for the two-body system).  For this reason, we do not use
phenomenological gradient corrections for a more accurate
determination of the constants $c_1$ and $c_2$.

Let us also investigate the deep bound-state limit ($\varepsilon
\rightarrow - \infty$ ) of the two-body system corresponding to a
positive scattering length $a\ll \lambda$ and zero range.  Taking this
limit in \Eq{master}, and noting that $\Gamma( x + 1/2)/\Gamma(x)
\rightarrow \sqrt{x}$ for $x \rightarrow \infty$, we find that the
binding energy is $\varepsilon \hbar \omega = -\hbar^2/(m a^2)$.
Thus, one can trivially write down the density functional for the
system in this limit as
\beq
\label{secondorder1}
{\cal E}_{B}[\rho] = -{\hbar^2\over 2 m a^2}\,\rho,
\eeq
and the energy per particle is $-{\hbar^2\over 2 m
a^2}$. Interestingly, this value coincides exactly with the $1/a^2$
correction that Bulgac and Bertsch~\cite{BB05} obtained from a fit to
Monte Carlo results close to the unitary regime, and it is about 20\%
larger than the analytical result that can be inferred from Steele's
work~\cite{Ste00}.

\begin{figure}
\centerline{\includegraphics*[width=6cm,angle=-90]{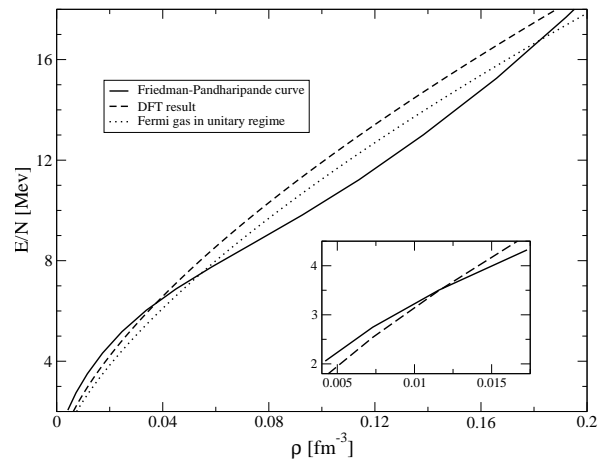}}
\vspace*{-.1in}
\caption{Energy per particle for symmetric neutron matter as a
function of the density. Full line:
Friedman-Pandharipande~\cite{FP81}; dashed line: result from DFT;
dotted curve: Fermi gas in unitary regime. The inset shows the DFT result
and includes finite range corrections.}
\label{fig:ks3}
\end{figure}

Finally, we apply \Eq{lda} to neutron matter, for which $a = -18.3$ fm
and $r_0 = 2.7$ fm. We drop the $r_0\rho^{1/3}$ term in
Eq.~(\ref{lda}), as this correction is only small for very small
densities.  In Fig.~\ref{fig:ks3} we compare our results to the
equation of state (EOS) by Friedman-Pandharipande~\cite{FP81}. Note
that that EOS is based on a realistic Hamiltonian, which includes
higher partial waves and three-body interactions. Recall that our
approach is limited to s-waves and two-body interaction. The inset of
Fig.~\ref{fig:ks3} shows the comparison for very small densities;
here, the correction due to the effective range is included, and the
restriction to $s$-waves is justified. We note that the inclusion
of the effective range correction for values of $r_0\rho^{1/3}$ less than 
$0.6$ improves the DFT result.

To summarize, we have considered interacting dilute Fermi systems near
the unitary regime and computed the corrections to its energy density
due to a large scattering length and a finite effective range of the
two-body interaction. Our calculations are based on the universality
of the density functional, and we determine its local density
approximation through comparison with exact results for the
harmonically trapped two-fermion system. The correction due to the
large scattering length agrees well with results from Monte Carlo
calculations and effective field theory, while the correction due to
the finite range implies a small systematic correction of order 0.01
to the universal constant extracted from Monte Carlo results. The
phenomenological inclusion of gradient terms is difficult due to finite-size
corrections. We also applied our results to neutron matter.

We are grateful to G.~E.~Astrakharchik, J. Carlson, S. Giorgini, and
S.~Y.~Chang for providing us with their data, and for discussions.  We
also thank A.~Bulgac, J.~Drut, R.~J.~Furnstahl, and D.~Lee for discussions.
This work was supported in part by the U.S. Department of Energy under
Contract Nos. DE-FG02-96ER40963 (University of Tennessee) and
DE-AC05-00OR22725 with UT-Battelle, LLC (Oak Ridge National
Laboratory).



\begin{thebibliography}{99} 

%
\bibitem{DeM99}
B. DeMarco and D. S. Jin,
Science {\bf 285}, 1703 (1999).
%
\bibitem{OHa02}
K. M. O'Hara, S. L. Hemmer, M. E. Gehm, S. R. Granade, and J. E. Thomas,
Science {\bf 298}, 2179 (2002).
%
\bibitem{Reg03}
C. A. Regal, C. Ticknor, J. L. Bohn, and D. S. Jin,
Nature {\bf 424}, 47 (2003), cond-mat/0305028.
%
\bibitem{Bou03}
T. Bourdel, J. Cubizolles, L. Khaykovich, K. M. F. Magalh{\~a}es,
S. J. J. M. F. Kokkelmans, G. V. Shlyapnikov, and C. Salomon,
\prl {\bf 91}, 020402 (2003), cond-mat/0303079.
%
\bibitem{Zwi03}
M. W. Zwierlein, C. A. Stan, C. H. Schunck, S. M. F. Raupach, S. Gupta,
Z. Hadzibabic, and W. Ketterle,
\prl {\bf 91}, 250401 (2003), cond-mat/0311617.
%
\bibitem{Geh03}
M. E. Gehm, S. L. Hemmer, S. R. Granade, K. M. O'Hara, and J. E. Thomas,
\pra {\bf 68}, 011401(R) (2003), cond-mat/0212499.
%
\bibitem{Reg04}
C. A. Regal, M. Greiner, and D. S. Jin,
\prl {\bf 92}, 040403 (2004), cond-mat/0401554.
%
\bibitem{Kin04}
J. Kinast, S. L. Hemmer, M. E. Gehm, A. Turlapov, and J. E. Thomas,
\prl {\bf 92}, 150402 (2004), cond-mat/0403540.
%
\bibitem{Bar04}
M. Bartenstein, A. Altmeyer, S. Riedl, S. Jochim, C. Chin,
J. Hecker Denschlag, and R. Grimm,
\prl {\bf 92}, 203201 (2004), cond-mat/0403716.
%
\bibitem{Bou04}
T. Bourdel, L. Khaykovich, J. Cubizolles, J. Zhang, F. Chevy, M. Teichmann,
L. Tarruell, S. J. J. M. F. Kokkelmans, and C. Salomon,
\prl {\bf 93}, 050401 (2004), cond-mat/0403091.
%
\bibitem{Gre05}
M. Greiner, C. A. Regal, and D. S. Jin,
\prl {\bf 94}, 070403 (2005), cond-mat/0407381.
%
\bibitem{Kin05}
J. Kinast, A. Turlapov, J. E. Thomas, Q. Chen, J. Stajic, and K.
Levin, Science {\bf 307}, 1296 (2005).
%
\bibitem{Hol01}
M. Holland, S. J. J. M. F. Kokkelmans, M. L. Chiofalo, and R. Walser,
\prl {\bf 87}, 120406 (2001), cond-mat/0103479.
%
\bibitem{Men02}
C. Menotti, P. Pedri, and S. Stringari,
\prl {\bf 89}, 250402 (2002), cond-mat/0208150.
%
\bibitem{Bul03}
A. Bulgac and Y. Yu,
\prl {\bf 91}, 190404 (2003), cond-mat/0303235.
%
\bibitem{Kim04}
Y. E. Kim and A. L. Zubarev,
\pra {\bf 70}, 033612 (2004), cond-mat/0404513.
%
\bibitem{Bul05}
A. Bulgac and G. F. Bertsch,
\prl {\bf 94}, 070401 (2005), cond-mat/0404687.
%
\bibitem{Ho03}
Tin-Lun Ho,
\prl {\bf 92}, 090402 (2004), cond-mat/0309109.
%
\bibitem{Hei01}
H. Heiselberg,
\pra {\bf 63}, 043606 (2001), cond-mat/0002056.
%
\bibitem{Sch05}
A. Schwenk and C. J. Pethick,
\prl {\bf 95}, 160401 (2005), nucl-th/0506042.
%

\bibitem{Car03}
J. Carlson, S.-Y. Chang, V. R. Pandharipande, and K. E. Schmidt,
\prl {\bf 91}, 050401 (2003), physics/0303094.
%
\bibitem{Ast04}
G. E. Astrakharchik, J. Boronat, J. Casulleras, and S. Giorgini,
\prl {\bf 93}, 200404 (2004), cond-mat/0406113;
G.~E.\ Astrakharchik,  R.\ Combescot, X.\ Leyronas,
and S.\ Stringari, 
\prl {\bf 95}, 030404 (2005), cond-mat/050361.
%
\bibitem{Bul05b}
A. Bulgac, J. E. Drut, and P. Magierski,
cond-mat/0505374.
%
\bibitem{Ste00}
J. V. Steele,
nucl-th/0010066.
%
\bibitem{HK64}
P.\ Hohenberg and W.\ Kohn, 
Phys.\ Rev. {\bf 136}, B864 (1964).
%
\bibitem{KOHN65} 
W.\ Kohn and L.~J.\ Sham, 
Phys.\ Rev.\ A {\bf 140}, 1133 (1965).
%
\bibitem{Pap05}
T.\ Papenbrock, 
\pra {\bf 72}, 041603(R) (2005), cond-mat/0507183.
%
\bibitem{Bak99}
G. A. Baker,
\prc {\bf 60}, 054311 (1999).
%
\bibitem{Eng97}
J. R. Engelbrecht, M. Randeria, and C. A. R. S{\'a}de Melo,
\prb {\bf 55}, 15153 (1997).
%
\bibitem{Lee05}
Dean Lee,
cond-mat/0511332.
%
\bibitem{Cha04}
S. Y. Chang, J. Carlson, V. R. Pandharipande, and K. E. Schmidt,
\pra {\bf 70}, 043602 (2004), physics/0404115.
%
\bibitem{BB05}
A.\ Bulgac and G.~F.\ Bertsch, 
\prl {\bf 94}, 070401 (2005). 
%
\bibitem{Bus98}
T. Busch, B. Englert, K. Rzazewski, and M. Wilkens,
Found. Phys. {\bf 28}, 549 (1998).
%
\bibitem{Block}M.\ Block and M.\ Holthaus, 
Phys.\ Rev. \ A {\bf 65}, 052102 (2002). 
%
\bibitem{Abramowitz}M.\ Abramowitz and I.~A. \ Stegun, 
{\it Handbook of Mathematical Functions}, (Dover 1972), Ch.~19.
%
\bibitem{Huang}
K.\ Huang and C.~N.\ Yang,
Phys.\ Rev. {\bf 105}, 767 (1957).
%
\bibitem{Lenz}W.\ Lenz, 
Z. Phys. {\bf 56}, 778 (1929).

\bibitem{HAmmER00}
H.-W.\ Hammer and R.~J.\ Furnstahl,
Nucl.\ Phys.\ A {\bf 678}, 227 (2000).

\bibitem{FURNSTAHL04b}A.\ Bhattacharyya and R.~J.\ Furnstahl, 
 Nucl.\ Phys.\ A {\bf 747}, 268 (2005), nucl-th/0408014.  

\bibitem{AB3}
A.\ Bhattacharyya and R.~J.\ Furnstahl,
 Phys.\ Lett. \ B {\bf 607}, 259 (2005).


\bibitem{FP81}B.\ Friedman and V.~R.\ Pandharipande, 
 Nucl.\ Phys.\ A {\bf 361}, 502 (1981).
  


\end{thebibliography}
\end{document}